\documentclass[12pt]{article}
\usepackage{amsfonts}
\usepackage{latexsym}
\usepackage{amsmath}
\usepackage{amssymb}
\usepackage{amssymb}

\newcommand{\newsection}{    
\setcounter{equation}{0}\section}
\def\appendix#1{\addtocounter{section}{1}\setcounter{equation}{0}
\renewcommand{\thesection}{\Alph{section}}
\section*{Appendix \thesection\protect\indent \parbox[t]{11.15cm}{#1}}
\addcontentsline{toc}{section}{Appendix \thesection\ \ \ #1}}

\newcommand{\be}{\begin{eqnarray}}
\newcommand{\ee}{\end{eqnarray}}
\newcommand{\bea}{\begin{eqnarray}}
\newcommand{\eea}{\end{eqnarray}}
\newcommand{\ba}{\begin{array}}
\newcommand{\ea}{\end{array}}
\newcommand{\nn}{\nonumber \\}

\newenvironment{frcseries}{\fontfamily{frc}\selectfont}{}

\def \la {\label}

\def\cL{{\cal L}}

\font\mybb=msbm10 at 11pt

\def\bb#1{\hbox{\mybb#1}}

\def\bR {\bb{R}}

\def\tn {{\tilde{\nabla}}}

\begin{document}
\begin{titlepage}
\begin{center}
\vspace*{-1.0cm}

\vspace{2.0cm} {\Large \bf AdS  backgrounds from black hole horizons } \\[.2cm]

\vspace{1.5cm}
 {\large  U. Gran$^1$, J. Gutowski$^2$ and  G. Papadopoulos$^2$}

\vspace{1.0cm}

${}^1$ Fundamental Physics\\
Chalmers University of Technology\\
SE-412 96 G\"oteborg, Sweden\\

\vspace{0.5cm}
${}^2$ Department of Mathematics\\
King's College London\\
Strand\\
London WC2R 2LS, UK\\

\vspace{0.5cm}

\end{center}

\vskip 1.5 cm
\begin{abstract}

We utilize the classification of IIB horizons with 5-form flux to present
a unified description for the geometry of  $AdS_n$, $n=3,5,7$  solutions. In particular, we show that all such backgrounds can be constructed from 8-dimensional  2-strong Calabi-Yau geometries with torsion which admit some additional isometries.  We explore the geometry of  $AdS_3$ and $AdS_5$ solutions but we do not find  $AdS_7$ solutions.

\end{abstract}

\end{titlepage}



\section{Introduction}

The gravitational duals of gauge/gravity correspondences
\cite{maldacena} and flux compactifications \cite{duff} are (warped)
products of AdS and Minkowski spaces with some other ``internal''
manifold. Because of this, they have been the focus of intensive
investigations in the literature. Most of the work has been done on
supersymmetric solutions. More recently the attention  has been shifted
to a systematic construction of such solutions and   several
approaches have been proposed to find such solutions. One is to make
an ansatz for the fields and Killing spinors that respect the
symmetries of the solution, for a review see \cite{grana} and
references within. However, such choices   are  special and it is
not apparent that all solutions can be described in this way. In
\cite{kim2005} an alternative approach has been used which again
involves the use of an ansatz for the fields  but now combined with
the spinor bilinears technique  for solving the KSEs \cite{hullgut}.
This avoids the problem of making a special choice for the Killing
spinors. However because of  the complexity of solving the KSEs in
this way, the fluxes of some backgrounds  are not always chosen to
be the most general ones allowed by the symmetries, see
e.g.~\cite{kim2006}.  So again  the most general solutions may not have
been described. Moreover, all investigations so far have been
done on a case by case basis and no overall picture has emerged for
the geometry of all $AdS_n$ backgrounds independently of $n$.

It is advantageous to remove all the assumptions made for constructing
gravitational duals of gauge/gravity correspondences and flux compactification
backgrounds, and at the same time find a way to describe all solutions in a uniform way.
In this paper, we shall focus on the second question. For this, we shall use a straightforward coordinate
transformation, which is described in appendix A, to bring the metric of
 all (warped) products of AdS and
Minkowski spaces with an ``internal'' manifold $X$  to a form which is included in
the standard near horizon geometry of extreme black holes \cite{wald}. Then we shall utilize
the classification results for supersymmetric near horizon geometries in 10- and 11-dimensional supergravity theories \cite{hh, iibhor, mhor},
based on the spinorial technique for solving Killing spinor equations \cite{uggp}, to give a unified
description for the geometry of all these warped products.

The classification of near horizon geometries is centered around the description
of near horizon spatial sections ${\cal S}$. The ``near horizon section'' for the wrapped product  $AdS_n\times_w X$ is ${\cal S}=H_{n-2}\times_w X$,
 where $H$ is a hyperbolic space, while for $\bR^{n-1,1}\times_w X$ is
 ${\cal S}=\bR^{n-2}\times_w X$.  As a consequence the geometry of
 all $N\times_w X$, $N=\bR^{n-2}, H_{n-2} $, and in particular that of $X$, can be
 determined as a special case of the geometry of spatial  horizon  sections.
 Thus the classification of gravitational duals and vacua of compactifications
  with fluxes can be viewed locally as a special case of that of near horizon
  geometries for black holes.

  The examples which we shall present in detail are the AdS solutions in IIB supergravity. In particular, we shall utilize
the classification of the near horizon geometries of IIB supergravity \cite{iibhor} with only
5-form flux, see also \cite{iibfive},
to present a unified description of $M\times_w X$, $M=AdS_n, \bR^{n-1,1}$  geometries in this theory.
We shall mostly focus on $M=AdS_n$. This is because the $\bR^{n-1,1}\times_wX$
spacetimes can be viewed as a special case of $AdS_n\times_w X$ in the limit where
the radius of $AdS$ goes to infinity. In particular, we shall show that the spacetimes
with $AdS_3\times_w X^7$ and $AdS_5\times_w X^5$  can be constructed from
8-dimensional
2-strong Calabi-Yau manifolds with torsion, i.e.~2-SCYT manifolds. This is the geometry of spatial sections of IIB horizons \cite{iibhor}. In addition, we shall demonstrate that there are no  $AdS_7\times_w X$ solutions in this class.

Furthermore, our  construction allows for $AdS_n\times_w X^{10-n}$ backgrounds for which the $SO(n-1,2)$ isometry group
of the metric is broken by the 5-form flux\footnote{The symmetry
preserved by the background does not contain an  $SO(n-1,1)$ subgroup and so a gauge/gravity duality interpretation of
such backgrounds is not apparent.}. To restore the full
$SO(n-1, 2)$ symmetry for the whole background, one has to impose additional restrictions on the geometry which we investigate.
It turns out that for the $AdS_3$ and $AdS_5$ backgrounds
which have full $SO(n-1, 2)$ symmetry, the geometry of $X$ coincides with that which has already been found in the literature.
In particular,  $X^5$ is Sasaki-Einstein for $AdS_5$
backgrounds \cite{se},  and $X^7$ is a fibration over a K\"ahler manifold for the $AdS_3$ backgrounds \cite{kim2005} for which
the Ricci scalar and Ricci tensor satisfy a certain
differential equation. Thus we establish a relation between 5-dimensional Sasaki-Einstein geometry
and the geometry on 6-dimensional K\"ahler manifolds of  \cite{kim2005} with the 2-SCYT geometry on 8-manifolds. We point out
though that in the latter case, the relationship
between the above K\"ahler geometry and the 2-SCYT geometry which arises from the horizon analysis \cite{iibhor} is
somewhat involved. The two differential systems which characterize the geometries are rather distinct.

The other examples which we shall explore are $AdS\times_w X$ solutions in heterotic supergravity. A direct inspection of the
classification results for supersymmetric solutions
in \cite{hetclas} and that of near horizon geometries \cite{hh} reveals
that there are two classes with one class consisting  of  non-trivial $SL(2,\bR)$ fibrations over a 7-dimensional base manifold.
Therefore the geometry of $AdS_3$ backgrounds can be
read off from that of heterotic horizons. The non-trivial fibration breaks the $SO(2,2)$
isometry group of $AdS_3$ to a subgroup which always contains one of the $SL(2,\bR)$ subgroups.   We describe
the geometry of the $AdS_3$ solutions which exhibit the full  $SO(2,2)$ symmetry.
We find that in this case the fibration is trivial and the solutions are direct products
$AdS_3\times X^7$. Depending on the geometry of $X^7$, the $AdS_3$ backgrounds which arise
 as a special case of heterotic horizons preserve  2, 4, 6
and 8 supersymmetries. This is in agreement with the results of \cite{ohta}.

Before we proceed, we would like to point out that the classification of near horizon geometries is carried out under two assumptions. One is the identification of the timelike or null Killing vector field of the near horizon geometry with the timelike or null Killing spinor bilinear of a supersymmetric background. The other is that the horizon section is compact which is
instrumental to solving the field equations. Neither of these two assumptions necessarily
hold in the investigation of $AdS\times_w X$ solutions. The first assumption on the
identification of the two Killing vector fields is technical and it can be removed, see
\cite{fourhor} for the near horizon geometries in 4-dimensional supergravity. The compactness
of the horizon section does not apply in the $AdS\times_w X$ case  if $X$ is not compact. However, the compactness of ${\cal S}$ is not always used in the analysis of near horizon solutions. For example
although it has been instrumental in the understanding of heterotic horizons in \cite{hh},
it has not been used in the analysis of IIB horizons in \cite{iibhor} admitting a pure Killing spinor. As a result our examples
of $AdS\times_wX$ spaces based on  IIB supergravity horizons are not restricted by the compactness assumption.

This paper has been organized as follows. In section two, we state the geometry of IIB horizons adapted for the investigation of $AdS$ backgrounds. In sections 3, 4 and 5,
we examine the geometry of solutions which are direct products of $AdS_3$, $AdS_5$ and $AdS_7$
spaces. In sections 6, 7 and 8 we examine the geometry of solutions which are warped products of $AdS_3$, $AdS_5$ and $AdS_7$ spaces, respectively. In particular in section 6, we establish
the relation between 2-SCYT geometry and the K\"ahler geometry of \cite{kim2005}.
In appendix A, we present the coordinate transformation which relates
$AdS$ solutions with near horizon geometries, and in appendix B, we describe the
$AdS_3$ backgrounds of heterotic supergravity.

\newsection{IIB CFT gravitational duals}

\subsection{IIB near horizon geometry}

The near horizon black hole geometries of IIB supergravity with only 5-form flux have been classified in \cite{iibhor}. The spatial horizon section ${\cal S}$ is an 8-dimensional manifold equipped with a 2-strong Calabi-Yau structure with torsion. This means that ${\cal S}$ is a Hermitian manifold, with Hermitian form $\omega$, such that
\bea
\hat\rho=0~,~~~~d(\omega\wedge H)\equiv \partial\bar\partial\omega^2=0~,
\la{2scyt}
\eea
where $\hat\rho$ is the Ricci form of the connection with skew-symmetric torsion
$H=-i (\partial-\bar\partial)\omega$.

The spacetime geometry can be summarized as follows. The spacetime metric and 5-form flux can be written as
\bea
ds^2&=&2 du (dr+ r h)+ ds^2({\cal S})
\cr
F &=& e^+ \wedge e^- \wedge Y - \star_8 Y ~,
\la{fields}
\eea
where
\bea
Y = \frac{1}{4}\left(d\omega - \theta_\omega \wedge \omega\right)~,
\eea
\bea
h=\theta_\omega
\la{lee}
\eea
and $\theta_\omega$ is  the
Lee form of ${\cal S}$. From now on instead of stating $F$, we shall give $Y$
for simplicity.

Therefore to construct solutions,
one has to find Hermitian manifolds ${\cal S}$ which satisfy the conditions
(\ref{2scyt}). Note that the Ricci form in (\ref{2scyt})   can be expressed as
\bea
\hat\rho=-i\partial\bar\partial \log \det g-di_I\theta_\omega~,
\la{hatr}
\eea
where $g$ is the Hermitian metric of ${\cal S}$.

\subsubsection{Geometry of gauge/gravity duals}

 Consider first the direct product  $AdS_n\times X^{10-n}$ backgrounds. The straightforward transformation which brings the metric of the above background
 into a near horizon form is explained in appendix A. One also finds that the
   spatial horizon section ${\cal S}$   is the direct product of a hyperbolic space with $X^{10-n}$, ${\cal S}=H_{n-2}\times X^{10-n}$, and the metric can be written as
\bea
ds^2({\cal S})=(Z^1)^2+\sum_k (Z^k)^2+ ds^2(X^{10-n})~.
\eea
To continue, we must identify the rest of the geometric structure on ${\cal S}=H_{n-2}\times X^{10-n}$ and in particular the Hermitian form $\omega$. There is not a natural way to do this. However,
suppose in addition that  $AdS_n$  is odd dimensional. In such a case $X^{10-n}$ is also odd-dimensional and so it admits a no-where vanishing vector field.
After possibly multiplying this vector field with a function of $X^{10-n}$, the metric on $X^{10-n}$ can be written as
\bea
ds^2(X^{10-n})= w^2+ ds^2_{(9-n)}~,
\la{metrn}
\eea
where $w$ is the 1-form dual to the vector field and $ds^2_{(9-n)}$ is orthogonal to $w$. In this case a Hermitian form can be defined on ${\cal S}$ as
\bea
\omega= Z^1\wedge w-\sum_k Z^{2k}\wedge Z^{2k+1}+ \omega_{(9-n)}~,
\la{herm}
\eea
where $\omega_{(9-n)}$ is a non-degenerate 2-form on $X$ in the directions transverse to $w$,
\bea
i_w \omega_{(9-n)}=0~.
\eea
The integrability of the associated almost complex structure is equivalent to requiring that the cylinder $\bR\times X^{10-n}$ is a complex manifold. Some of the integrability
conditions are
\bea
dw^{2,0}=0~,~~~i_w dw=0~,
\la{incon}
\eea
i.e.~$dw$ is (1,1)-form and $dw$ is transverse to $w$. Note that by definition $|w|^2=1$.

Furthermore in the direct product case comparing the expression for the metric of $AdS_n\times X^{10-n}$ in (\ref{adsn}) and (\ref{lee}),  one
has to take
\bea
\theta_\omega=-{2\over \ell} Z^1~,
\la{leecon}
\eea
which restricts the choice of Hermitian structure on ${\cal S}$ which in turn
restricts $X^{10-n}$. In addition to the integrability of the almost complex structure
on $H_{n-2}\times X^{10-n}$ and (\ref{leecon}), the two conditions in (\ref{2scyt}) must
also be satisfied. These conditions
will be investigated for each case separately.

In the warped product case, the geometric data are altered as follows. The metric on ${\cal S}$ is chosen as
\bea
ds^2({\cal S})= A^2 [(Z^1)^2+\sum_k (Z^k)^2]+ ds^2(X^{10-n})~,~~
\eea
where now
\bea
ds^2(X^{10-n})= A^2 w^2+ ds^2_{(9-n)}~,
\eea
where $A$ is the warp factor.
The Hermitian form is chosen as
\bea
\omega=A^2[ Z^1\wedge w-\sum_k Z^{2k}\wedge Z^{2k+1}]+ \omega_{(9-n)}~.
\la{wherm}
\eea
The integrability conditions of the associated almost complex structure
again imply (\ref{incon}).  Again comparing the expression for the warped product metric in (\ref{wpm}) and (\ref{lee}), one finds that the condition on the Lee form is now modified to
\bea
\theta_\omega=-{2\over \ell} Z^1-d\log A^2~.
\la{wlee}
\eea
As in the previous case, the two conditions (\ref{2scyt}) of the 2-SCYT structure should also
be satisfied and will be investigated separately for each case.

\newsection{Direct product $\mathrm{\bold{AdS}}_3$ gravitational duals}

The metric on the direct product $AdS_3\times X^7$ takes the form
\bea
ds^2_{(10)}=2( du dr -{2r \over \ell} du Z^1) + (Z^1)^2+ ds^2(X^7)~,~~~dZ^1=0~,
\eea
and so $Z^1=dx$.  Note that $AdS_3$
is spanned by the coordinates $x,u,r$. Moreover ${\cal S}_{(8)}=\bR\times X^7$.
Thus to find solutions, one has to find those 7-dimensional manifolds
$X^7$ such that the cylinder $\bR\times X^7$ admits a 2-SCYT structure.

As in the general case described in the previous section, we set
$ds^2(X^7)= w^2+ ds^2_{(6)}$,
where $ds^2_{(6)}$ is a metric transverse to $w$. We also  postulate the Hermitian
form
\bea
\omega_{(8)} =   Z^1 \wedge w  + \omega_{(6)}~,
\la{3herm}
\eea
 with $i_w \omega_{(6)}=0$, where
the subscripts in $\omega$ denote the dimension of the associated space.
The integrability of the almost complex structure implies (\ref{incon}).

The torsion 3-form is
\bea
H=w\wedge dw+ Z^1\wedge \xi- i_{I_{(4)}} \eta~,
\eea
where we have decomposed
\bea
d\omega_{(6)}= w\wedge \xi+ \eta~,~~~i_w\eta=0~.
\eea
Moreover, $\xi$ is (1,1), and $\eta$ is (2,1)+(1,2), with respect to $I_{(6)}$ as required by the integrability
of the complex structure $I_{(8)}$.

Next consider the condition (\ref{leecon}), $\theta_{\omega_{(8)}}=-{2\over \ell} Z^1$, to find\footnote{Let $\rho$ be a k-form, then $\rho\cdot \omega$ is
the un-weighted contraction of the first two indices of $\rho$ with the indices
of $\omega$.} that
\bea
dw\cdot \omega_{(6)}={4\over \ell}~,~~~\xi\cdot \omega_{(6)}=0~,~~~\eta\cdot \omega_{(6)}=0~.
\la{3lee}
\eea

Moreover,  the 2-SKT condition (\ref{2scyt}), $d(\omega_{(8)}\wedge H)=0$, implies that
\bea
w\wedge di_{I_{(6)}} \eta+w\wedge \xi\wedge \xi+ \eta\wedge \xi+\omega_{(6)}\wedge d\xi=0~,~~~dw\wedge \eta=0~,~~~\omega_{(6)}\wedge dw\wedge dw=0~.
\nonumber \\
\la{32skt}
\eea
The first condition can be simplified a bit further by writing
\bea
&&di_{I_{(6)}}\eta= w\wedge \alpha+\beta~,~~~i_w\beta=0~,
\cr
&&d\xi=w\wedge \gamma+ \delta~,~~~i_w\delta=0~,
\eea
to find
\bea
\beta+\xi\wedge \xi+\omega_{(6)}\wedge \gamma=0~,~~~\eta\wedge \xi+\omega_{(6)}\wedge \delta=0~.
\eea

It remains to solve the first equation in (\ref{2scyt}),  $\hat\rho=0$. This has been expressed
in (\ref{hatr})
in terms of the determinant of the Hermitian metric and the Lee form of the 2-SCYT manifold
${\cal S}$. A straightforward
computation reveals that this condition can be rewritten in terms of geometric data on $X^7$ as
\bea
&&-{w^i\over w^2} \partial_i\log\det g_{(7)}+ \Lambda=0~,
\cr
&&{1\over 4} di_{I_{(6)}} d \log\det g_{(7)}+{2\over \ell} dw=0~.
\la{3rho}
\eea
where $\Lambda$ is constant and $g_{(7)}$ is the Riemannian metric on $X^7$.
The last equation requires that
\bea
{\cal L}_wi_{I_{(6)}} d \log\det g_{(7)}=0~.
\eea

To summarize, the metric and the flux of the solutions can be written as
\bea
ds^2_{(10)}&=&2( du dr -{2r \over \ell} du Z^1) + (Z^1)^2+  w^2+ ds^2_{(6)}~,~~~dZ^1=0~,
\cr
Y &=& \frac{1}{4}\left( Z^1 \wedge (-dw+ \frac{2}{l} \omega_{(6)}) + w \wedge \xi + \eta \right) ~,
\la{3exfields}
\eea
where the 5-form flux $F$ is given in terms of $Y$ as in (\ref{fields}). The 2-SCYT
structure on the horizon section ${\cal S}=\bR\times X^7$ is associated with the Hermitian form (\ref{3herm}) and the  geometry
is constrained by (\ref{3lee}), (\ref{32skt}) and (\ref{3rho}).

\subsection{$SO(2,2)$ invariant backgrounds}

It is apparent from (\ref{3exfields}) that the 5-form flux breaks the $SO(2,2)$ isometry of $AdS_3$ space.  This is unless
\bea
\xi=\eta=0~.
\eea
In this case, the conditions on the geometry reduce to
\bea
d\omega_{(6)}=0~,~~~dw\cdot \omega_{(6)}={4\over \ell}~,~~~\omega_{(6)}\wedge dw\wedge dw=0~,
\la{sym3ads}
\eea
and (\ref{3rho}). As we shall demonstrate later for warped products, which includes
the direct product case presented here, the $SO(2,2)$ symmetric backgrounds are the same as those presented in \cite{kim2005}. However, the way that the two different descriptions
of the geometry of $AdS_3$ backgrounds are related is non-trivial.

\subsection{Examples}

Although the description of the geometry on $\bR\times X^7$ is simple, the conditions
on $X^7$ that arise from implementing the geometric restrictions are rather involved.
Moreover, one can easily see that  apparent  geometries on $X^7$, like for example (Einstein)
 Sasakian,  are not solutions. To give some examples, we shall focus on the solutions with
 $SO(2,2)$ isometry and take that $w$ generates a holomorphic isometry\footnote{This is not an assumption as we shall demonstrate in section 6.1.} in $\bR\times X^7$. Moreover, let us take $X^6$ to be a product of K\"ahler-Einstein
spaces. Since $X^6$ is 6-dimensional it can be the product of up to 3 such spaces.
Writing the Ricci form of these spaces as $\rho_i=-\lambda_i \omega_i$, where $\lambda_i$ and
$\omega_i$
is the cosmological constant and the K\"ahler form of each subspace, respectively,
and taking $\omega_{(6)}=\sum_i \omega_i$, we have that
\bea
dw={\ell\over2}\sum_i \lambda_i \omega_i~.
\eea
Then the second condition in (\ref{sym3ads}) gives
\bea
\sum_i \lambda_i n_i={8\over \ell^2}
\eea
and the third condition in (\ref{sym3ads}) implies either
\bea
\lambda_1 \lambda_2 + \lambda_1 \lambda_3 + \lambda_2 \lambda_3 =0
\eea
if $n_1=n_2=n_3=2$, or
\bea
2 \lambda_1 \lambda_2 + \lambda_2^2 =0
\eea
if $n_1=2, n_2=4$;
where $n_i$ is the real dimension of the i-th subspace of $X^6$. There are solutions
to these conditions by taking $X^6=S^2\times T^4$ and $X^6=S^2\times K_3$. Such examples of 2-SCYT geometries have been investigated in \cite{iibhor} and our result in section
6.1 relates them to those investigated in \cite{kim2}.

\newsection{Direct product $\mathrm{\bold{AdS}}_5$ gravitational duals}

The metric  is
\bea
ds^2_{(10)}=2du (  dr -{2r \over \ell} Z^1) + (Z^1)^2+(Z^2)^2+(Z^3)^2+ ds^2(X^5)
\eea
where  $Z^1, Z^2, Z^3$ are given in (\ref{zframe}) and satisfy the differential system
(\ref{dzet}). The $AdS_5$ subspace is spanned by the coordinates $u, r, z, x^k$.

Next
writing the metric on $X^5$ as in (\ref{metrn}),
the Hermitian form is chosen  as in (\ref{herm}), i.e.
\bea
\omega_{(8)} =   Z^1 \wedge w - Z^2 \wedge Z^3 + \omega_{(4)}~,
\la{5herm}
\eea
where $\omega_{(4)}$ is a 2-form on $X^5$ such that
$i_w \omega_{(4)}=0$.
  The integrability
of the complex structure implies (\ref{incon}).

The torsion 3-form of ${\cal S}_8$ is
\bea
H=w\wedge dw+{2\over \ell} w\wedge Z^2\wedge Z^3+ Z^1\wedge \xi- i_{I_{(4)}} \eta~,
\eea
where we have decomposed
\bea
d\omega_{(4)}= w\wedge \xi+ \eta~,~~~i_w\eta=0~.
\la{5xieta}
\eea
Moreover, $\xi$ is (1,1) with respect to $I_{(4)}$ as required by the integrability
of the complex structure $I_{(8)}$.

It remains to find the restrictions on the geometry of
$X^5$ imposed by  the condition on the Lee form (\ref{leecon}) and the  2-SCYT  condition
(\ref{2scyt}) on ${\cal S}_8$.  First, the Lee condition (\ref{leecon}) $
\theta_{\omega_{(8)}}=-{2\over \ell} Z^1$
 gives that
\bea
-{1\over2} dw\cdot \omega_{(4)} +{4\over \ell}=0~,~~~\xi\cdot \omega_{(4)}=0~,~~~\eta=0~.
\la{ads5lee}
\eea
Then, the 2-SKT condition $d(\omega_{(8)}\wedge H)=0$ gives
\bea
-w\wedge dw+{\ell\over2} d\xi+{2\over \ell} w\wedge \omega_{(4)}=0~,
\la{52skt}
\eea
where to derive the above condition we have used the results obtained  in (\ref{ads5lee}).

It remains to solve the 2-SCYT condition $\hat\rho=0$ (\ref{hatr}). A straightforward
computation reveals the conditions
\bea
&&-{w^i\over w^2} \partial_i\log\det g_{(5)}+ \Lambda=0~,
\cr
&&{1\over 4} di_{I_{(4)}} d \log\det g_{(5)}+{3\over \ell} dw=0~,
\la{5rho}
\eea
where $\Lambda$ is constant and $g_{(5)}$ is the Riemannian metric on $X^5$.
The last equation requires that
\bea
{\cal L}_wi_{I_{(4)}} d \log\det g_{(5)}=0~.
\eea

To summarize, the solutions can be written as
\bea
ds^2_{(10)}&=&2( du dr -{2r \over \ell} du Z^1) + \sum_{k=1}^3(Z^k)^2+  w^2+ ds^2_{(4)}~,~~~
\cr
Y &=&  \frac{1}{4}\left(  - \frac{4}{l}Z^1\wedge Z^2 \wedge Z^3 + Z^1 \wedge (-dw+ \frac{2}{l} \omega_{(4)})  + w \wedge \xi  \right) ~,
\la{5exfields}
\eea
where again the 5-form flux is given in terms of $Y$ as in (\ref{fields}). The 2-SCYT
structure on the near horizon section ${\cal S}= H_3\times X^5$  is with respect to the Hermitian form (\ref{5herm}) and the restrictions
on the geometry are given in (\ref{ads5lee}), (\ref{52skt}) and (\ref{5rho}).

\subsection{$SO(4,2)$ invariant backgrounds.}

It is clear that the 5-form flux is not invariant under the full $SO(4,2)$ symmetry
of the $AdS_5$ subspace unless
\bea
\xi=0~,~~~\omega_{(4)}={\ell\over 2} dw~.
\eea
As a result $X^5$ is a Sasakian manifold. The remaining conditions
imply that in addition $X^5$ is Einstein. Thus for $\xi=0$, $X^5$
is a Sasaki-Einstein manifold. This class of solutions is well known
 \cite{se} and include the $AdS_5\times S^5$ solution
of IIB supergravity.

\newsection{$\mathrm{\bold{AdS}}_7$}

The metric is
\bea
ds^2_{(10)}=2( du dr -{2r \over \ell} du Z^1) + (Z^1)^2+(Z^2)^2+(Z^3)^2+ (Z^4)^2
+(Z^5)^2+ds^2(X^3)
\eea
where $Z^1, Z^2, Z^3, Z^4, Z^5$ are given in (\ref{zframe}) and (\ref{dzet}).
The coordinates $(u,r, z, x^k)$ span $AdS_7$.

 Furthermore $X^3$ is an odd dimensional manifold and the metric can be written as
 in (\ref{metrn}), $ds^2(X^3)= w^2+ ds^2_{(2)}$
where $ds^2_{(2)}$ is  transverse to $w$. The Hermitian
form is
\begin{eqnarray}
\omega_{(8)} =   Z^1 \wedge w - Z^2 \wedge Z^3 -Z^4\wedge Z^5+ \omega_{(2)} \ ,
\end{eqnarray}
where $\omega_{(2)}$ is a 2-form on $X^3$ such that $i_w \omega_{(2)}=0$
and it is hermitian with respect to  $ds^2_{(2)}$. The integrability of the complex
structure implies (\ref{incon}). Though the condition $dw^{2,0}=0$ is automatic in this case.

Next the torsion 3-form can be easily computed to find
\bea
H=w\wedge dw+{2\over \ell} w\wedge (Z^2\wedge Z^3+Z^4\wedge Z^5)+ Z^1\wedge \xi
\eea
where we have written
\bea
d\omega_{(2)}= w\wedge \xi~,~~~i_w\xi=0~.
\eea
Observe that $\xi$ is (1,1),  and that the 3-form,  $\eta$,
 vanishes identically.

Next imposing the condition on the Lee form, $\theta_{\omega_{(8)}}=-{2\over \ell} Z^1$,
 one gets
\bea
-{1\over2} dw\cdot \omega_{(4)} +{6\over \ell}=0~,~~~\xi\cdot \omega_{(2)}=0~.
\eea
The latter condition implies that
\bea
\xi=0~.
\eea
Next take the 2-SKT condition $d(\omega_{(8)}\wedge H)=0$ to find
\bea
w=0~.
\eea
So there are no such solutions.

\newsection{Warped $\mathrm{\bold{AdS}}_3$}

Having explained the direct product case $AdS_n\times X^{10-n}$ in some detail, we shall only outline the
key points that arise in the derivation of the geometric conditions  for the warped products
$AdS_n\times_w X^{10-n}$ to be solutions of IIB supergravity.  In particular for the $AdS_3$ case, we have that the Hermitian form is
\bea
\omega= A^2 [ Z^1\wedge w]+ \omega_{(7)}~,~~~dZ^1=0~.
\la{w3herm}
\eea
To solve the conditions on the Lee form and those required by the 2-SCYT structure,
 we first compute the
skew-symmetric torsion 3-form to find
\bea
H=-A^2 i_{I_{(6)}} \rho \wedge Z^1\wedge w+ A^2 w\wedge dw+
Z^1\wedge \xi- i_{I_{(6)}}\eta~,
\eea
where we have used
\bea
d\omega_{(6)}=w\wedge \xi+ \eta~,~~~d\log A^2=f w+\rho~,~~~i_w\eta=i_w\rho=0~.
\la{frho}
\eea

The condition on the Lee form (\ref{wlee}) gives
\bea
{1\over2}\xi\cdot\omega_{(6)}=-f~,~~~{1\over2} A^2 dw\cdot \omega_{(6)}={2\over \ell}~,~~~{1\over2}i_{I_{(6)}}((i_{I_{(6)}}\eta)\cdot \omega_{(6)})=2\rho~.
\label{w3lee}
\eea

Next the 2-SKT condition  $d(\omega_{(8)}\wedge H)=0$ leads to
\bea
\label{w32skt}
&&A^2 dw\wedge dw\wedge \omega_{(6)}+\lambda \wedge \omega_{(6)}^2-
2 i_{I_{(6)}}\rho \wedge \eta\wedge \omega_{(6)}=0~,
\cr
&&-A^2 \rho\wedge dw\wedge \omega_{(6)}-A^2 dw\wedge \eta+\mu\wedge \omega_{(6)}^2+
2 i_{I_{(6)}}\rho\wedge \xi\wedge \omega_{(6)}=0~,
\cr
&&A^2 dw\wedge i_{I_{(6)}}\eta-\pi\wedge \omega_{(6)}-\xi\wedge \eta
+A^2 i_{I_{(6)}} \rho\wedge dw\wedge \omega_{(6)}=0~,
\cr
&&-w\wedge d\big( i_{I_{(6)}} \eta A^2\big)-w\wedge d\big( i_{I_{(6)}} \rho A^2)
\wedge \omega_{(6)}+A^2 w\wedge i_{I_{(6)}}\rho\wedge \eta
\cr
&&-w\wedge \xi^2-w\wedge\zeta\wedge \omega_{(6)}=0~,
\eea

where
\bea
d(i_{I_{(6)}} \rho)=w\wedge \mu+\lambda~,~~~d\xi=w\wedge \zeta+ \pi~,~~~d\eta=-dw\wedge
\xi+w\wedge \pi~.
\eea

It remains to solve the other 2-SCYT condition in (\ref{2scyt}), $\hat\rho=0$. A direct
substitution reveals that (\ref{hatr}) gives
\bea
-{1\over4}{w^i\over w^2} \partial_i\log\det g_{(7)}-{5\over4} f+\Lambda=0~,
\cr
{1\over4} di_{I_{(6)}}d\log\det g_{(7)}+{2\over\ell} dw+ {5\over4} di_{I_{(6)}}\rho=0~,
\la{w3rho}
\eea
where $\Lambda$ is constant.

To summarize, the fields are given by
\bea
ds^2&=&2 du (dr-{2r\over \ell} Z^1-rd\log A^2)+  A^2 (Z^1)^2+ ds^2(X^7)
\cr
Y&=&  \frac{1}{4}\left(  Z^1 \wedge (- A^2 dw+ 2 A^2 w\wedge\rho + \frac{2}{l} \omega_{(6)}) + w \wedge \xi + \eta  + (fw+\rho)\wedge \omega_{(6)} \right)~,
\la{w3fields}
\eea
where the 5-form flux is given in terms of $Y$ as in (\ref{fields}). The Hermitian
form on ${\cal S}=\bR\times X^7$ is given in (\ref{w3herm}) and the geometric conditions that restrict the geometry are given in (\ref{w3lee}), (\ref{w32skt}) and (\ref{w3rho}).

\subsection{Backgrounds with $SO(2,2)$ symmetry}

Requiring that the solution is invariant under the $SO(2,2)$ symmetry of $AdS_3$,
we shall demonstrate that $X^7$ is a fibration over a 6-dimensional K\"ahler manifold.
Clearly as in the direct product case, the 5-form flux is not invariant under $SO(2,2)$ isometries of the metric. Examining the expression for the 3-form $Y$ as given in (\ref{w3fields}),  one find that it is $SO(2,2)$ invariant provided that
\bea
\label{sym1}
\eta+ \rho \wedge \omega_{(6)}=0~,
\eea
and
\bea
\label{sym2}
\xi + f \omega_{(6)}=0~.
\eea
The first condition in ({\ref{w3lee}}) together with ({\ref{sym2}}) imply that
\bea
\xi=0, \qquad f=0~.
\la{connn}
\eea
In addition, the third condition in ({\ref{w3lee}}) is implied by ({\ref{sym1}}).
It is also straightforward to show that the third and fourth conditions obtained from the
2-SCYT condition ({\ref{w32skt}}) hold automatically, leaving only the first two conditions.
These can be written as
\bea
\label{tx0}
A^2 dw \wedge dw \wedge \omega_{(6)} + d \big( i_{I_{(6)}} \rho \big) \wedge \omega_{(6)}
\wedge \omega_{(6)} + 2  \big( i_{I_{(6)}} \rho \big) \wedge \rho \wedge \omega_{(6)} \wedge \omega_{(6)}
=0~.
\eea
Note in particular that
\bea
\mu=0 \ .
\eea
Also, ({\ref{tx0}}) can be rewritten as
\bea
\label{tx0b}
{\tilde{\nabla}}^2 \log A = A^{-2} \bigg( -{1 \over \ell^2} + {1 \over 8} d(A^2 w). d (A^2 w) \bigg)~,
\eea
where $\tn$ is the Levi-Civita connection on $X^7$.

To proceed, using the Einstein equation, we find that the Ricci tensor of $X^7$ is
\bea
\label{rtens1}
{\tilde{R}}_{ab}= {3 \over 2} \tn_a \tn_b \log A^2 +{3 \over 4} \tn_a \log A^2 \tn_b \log A^2
-4Y_{a ij} Y_b{}^{ij} +{2 \over 3} g_{ab} Y_{i_1 i_2 i_3} Y^{i_1 i_2 i_3}
\nn
\eea
where
\bea
Y = A Z^1 \wedge \bigg( w \wedge dA +{1 \over 2 \ell} A^{-1} \omega_{(6)} -{1 \over 4} A dw \bigg)~,
\eea
 $a,b$ are frame indices on $X^7$ and  $i, j$
are frame indices on ${\cal{S}}$. Note that
\bea
Y_{n_1 n_2 n_3} Y^{n_1 n_2 n_3} = 6 A^{-2} dA. dA +{3 \over 2 \ell^2} A^{-2} +{3 \over 16} A^2 (dw).(dw) \ .
\eea
Next, set
\bea
\kappa = A^2 w
\eea
and compute
\bea
\label{tx1}
\tn^2 \kappa^2 = 2 \tn^{(a} \kappa^{b)} \tn_{(a} \kappa_{b)}+{1 \over 2} d \kappa . d \kappa
+2 \kappa^b \tn^a (d \kappa)_{ab} +2 \kappa^b \tn^a \tn_b \kappa_a~.
\eea
The terms on the RHS of ({\ref{tx1}}) can be simplified further on using
\bea
\kappa^b \tn^a (d \kappa)_{ab} &=& - \tn^a (\kappa^b d \kappa_{ba}) -{1 \over 2} d \kappa .  d \kappa
\nn
&=& \tn^2 A^2 -{1 \over 2} d \kappa . d \kappa
\eea
and
\bea
2 \kappa^b \tn_a \tn_b \kappa^a = 2 \kappa^a \kappa^b {\tilde{R}}_{ab}
= -2 dA. dA +2 \ell^{-2} +{1 \over 4} A^4 dw. dw
\eea
where we note that
\bea
\kappa^a \kappa^b \tn_a \tn_b \log A^2 = 2 dA . dA \ .
\eea
The LHS of ({\ref{tx1}}) can also be rewritten in terms of $A$, using
\bea
\kappa^2 = A^2 \ .
\eea

Then, on comparing  ({\ref{tx0b}}) with ({\ref{tx1}}), one finds that
\bea
 \tn^{(a} \kappa^{b)} \tn_{(a} \kappa_{b)} =0
\eea
so $\kappa$ is an isometry of $X$, which also satisfies
\bea
\cL_\kappa A=0, \qquad \cL_\kappa w =0 \ .
\eea

We rewrite the metric on $X^7$ as
\bea
\label{metX}
ds^2(X^7) = A^2 w^2 + A^{-2} d\mathring s^2(B^6)
\eea
where $d\mathring s^2(B^6)$ does not depend on the coordinate along the isometry. It turns out
that $B^6$ equipped with $d\mathring s^2(B^6)$ and Hermitian form
\bea
\label{kahX}
\mathring\omega_{(6)} = A^2 \omega_{(6)}~,
\eea
is a K\"ahler manifold.  This follows from (\ref{sym1}), (\ref{sym2}) and (\ref{connn}).

Furthermore,  decomposing the Ricci tensor in ({\ref{rtens1}}) along the directions of $B^6$, one finds
\bea
\label{Bricform}
{\mathring {R}}_{\alpha \beta} = {2 \over \ell} (I_{(6)})^\gamma{}_\alpha (d w)_{\gamma\beta} \eea
where ${\mathring{R}}$ is the Ricci tensor of $B^6$ and $\alpha, \beta$ are (real) frame indices on $B^6$.
Thus Ricci form of $B^6$ constructed from the K\"ahler metric is proportional to $dw$.

In addition, the second condition in ({\ref{w3lee}}) gives
\bea
\label{tracesig}
(d w) \ . \  \mathring\omega_{(6)} = 4 \ell^{-1} A^{-4}~,
\eea
where now the contraction is taken with respect to the K\"ahler metric on $B^6$. This together with
({\ref{Bricform}}) implies that the Ricci scalar of $B^6$ is
\bea
\label{conffact}
{\mathring{R}} = {8 \over \ell^2} A^{-4}~.
\eea

Next, we return to ({\ref{tx0b}}). This can be rewritten in terms of the curvature of $B^6$
as
\be
\label{kimeq}
{\mathring{\nabla}}^2 {\mathring{R}} = {1 \over 2} {\mathring{R}}^2 - {\mathring{R}}_{\alpha \beta} {\mathring{R}}^{\alpha \beta}~.
\eea
This equation has been found before in the context of $AdS_3$ solutions
in IIB supergravity in {\cite{kim2005}}. Our result establishes a non-trivial
relationship between some 2-SCYT manifolds and K\"ahler manifolds\footnote{Not all 2-SCYT
manifolds are related to K\"ahler manifolds in this way.}. In particular, if ${\cal S}$ is an 8-dimensional 2-SCYT manifold with metric
\bea
ds^2({\cal S}) = A^2[ (Z^1)^2+ w^2]+ A^{-2} d\mathring s^2(B^6)~,
\eea
where $ Z^1, w$ generate commuting isometries, $dZ^1=0$,  and with Hermitian form given in (\ref{w3herm}),
then $B^6$ is a K\"ahler manifold with K\"ahler form given in (\ref{kahX}) whose Ricci data satisfy (\ref{kimeq}). Furthermore the curvature $dw$ of the fibration is proportional to the Ricci form of $ B^6$, i.e.~the fibration is the canonical fibration,
and the warp factor is determined in terms of the Ricci scalar. It is remarkable
that the 2-SCYT data quadratic in derivatives turn into an equation 4th order in
derivatives on the metric of $B^6$. A similar equation arises also
in 11 dimensions and in \cite{mhor} the question was raised whether there always
exists a solution on K\"ahler manifolds.

\newsection{Warped $\mathrm{\bold{AdS}}_5$  }

As in the warped $AdS_3$ case, the Hermitian form is
\bea
\omega_{(8)}= A^2(Z^1\wedge w- Z^2\wedge Z^3)+\omega_{(4)}~,
\la{w5harm}
\eea
which gives rise to the torsion 3-form
\bea
&&H=-A^2 f Z^1\wedge Z^2\wedge Z^3- A^2 i_{I_{(4)}} \rho \wedge (Z^1\wedge w-
Z^2\wedge Z^3)
\cr
&&+ A^2 w\wedge (dw+{2\over\ell} Z^2\wedge Z^3)+ Z^1\wedge \xi-i_{I_{(4)}} \eta~.
\eea
Using these, one finds that the Lee form condition (\ref{wlee})  gives
\bea
-{1\over2} A^2 dw\cdot \omega_{(4)}+{4\over\ell} =0~,~~~
{1\over2} \xi\cdot \omega_{(4)}=-2 f~,~~~
{1\over2} i_{I_{(4)}} (i_{I_{(4)}} \eta\cdot \omega_{(4)})=3\rho~,
\la{w5lee}
\eea
where $f$ and $\rho$  are defined as in (\ref{frho}).

It remains to investigate the two 2-SCYT conditions in (\ref{2scyt}). In particular,
 $d(\omega_{(8)}\wedge H)=0$ after some manipulation gives  that
\bea
d \bigg(2 A^4 w \wedge i_{I_{(4)}} \rho -A^2 \xi -f A^2 \omega_{(4)} \bigg)
+{2 \over \ell} A^4 w \wedge dw
\nonumber  \\
 +{4 \over \ell} A^2 \omega_{(4)} \wedge  i_{I_{(4)}} \rho
-{4 \over \ell^2} A^2 \omega_{(4)} \wedge w =0~,
\la{w52skt1}
\eea
and
\bea
d \bigg( \omega_{(4)} \wedge \big( -f \omega_{(4)} +
A^2 w \wedge  i_{I_{(4)}} \rho \big) \bigg)=0~,
\la{w52skt2}
\eea
where $\xi$ is defined as in the direct product case in (\ref{5xieta}).
Using  (\ref{hatr}), the other 2-SCYT condition (\ref{2scyt})  gives
\bea
-{1\over4}{w^i\over w^2} \partial_i\log\det g_{(5)}-{7\over4} f+\Lambda=0~,
\cr
{1\over4} di_{I_{(4)}}d\log\det g_{(5)}+{3\over\ell} dw+ {7\over4} di_{I_{(4)}}\rho=0~,
\la{w5rho}
\eea
where $\Lambda$ is constant.

To summarize, the fields for warped $AdS_5$ backgrounds are given by
\bea
ds^2&=&2 du (dr-{2r\over \ell} Z^1-rd\log A^2)+  A^2[\sum_{k=1}^3 (Z^k)^2]+ ds^2(X^5)~,
\cr
Y &=&  \frac{1}{4}\big( - \frac{4}{l} A^2 Z^1\wedge Z^2 \wedge Z^3  + Z^1 \wedge (- A^2 dw+ 2 A^2 w\wedge\rho + \frac{2}{l} \omega_{(4)})
\cr
 &&
  - 2A^2 Z^2\wedge Z^3 \wedge (fw+\rho)+ w \wedge \xi + \eta  + (fw+\rho)\wedge \omega_{(4)} \big).
  \la{w5fields}
\eea
where again the 5-form field strength is given in terms of $Y$ as in (\ref{fields}).
The underlying  geometry is 2-SCYT with hermitian form (\ref{w5harm}) and the
geometry is restricted as in (\ref{w5lee}), (\ref{w52skt1}), (\ref{w52skt2})
and (\ref{w5rho}).

\subsection{$SO(4,2)$ invariant solutions}

As in the direct product case, the flux given in (\ref{w5fields}) is not invariant
under the $SO(4,2)$ isometries of the metric. Enforcing $SO(4,2)$ symmetry for the whole background, we find that
\bea
f=\rho=\xi=\eta=0~.
\eea
This implies that  $A$ is constant. As a result, the warped product becomes direct
 and the analysis of the $SO(4,2)$ invariant solutions is identical to that given
 in the direct product case.

\newsection{Warped $\mathrm{\bold{AdS}}_7$}

As for direct product $AdS_7$ backgrounds, we shall show that there are no warped $AdS_7$ solutions. For this,
the Hermitian form and torsion 3-form are
\bea
\omega_{(8)}= A^2 \big(Z^1\wedge w-Z^2\wedge Z^3-Z^3\wedge Z^4\big)+ \omega_{(2)}~,
\eea
and
\bea
&&H=-A^2 f Z^1\wedge Z^2\wedge Z^3- fA^2 Z^1\wedge Z^4\wedge Z^5-  A^2\,i_{I_{(2)}}\rho
\wedge (Z^1\wedge w- Z^2\wedge Z^3-Z^4\wedge Z^5)
\cr
&&+ A^2  w\wedge (dw+{2\over \ell} Z^2\wedge Z^3+{2\over\ell} Z^4\wedge Z^5)
+ Z^1\wedge \xi~,
\eea
respectively.

Next the condition  on the Lee form (\ref{wlee}) gives
\bea
-{1\over2} A^2 dw\cdot \omega_{(2)}+{6\over\ell} =0~,
\cr
3 f+{1\over2} \xi\cdot \omega_{(2)}=0~,~~~
\rho=0~.
\la{wtheta7}
\eea
This in turn implies that
\bea
d\log A^2=f w~,
\eea
and so
\bea
df w+ f dw=0~.
\la{dfw}
\eea
Now if $f=0$, then $A$ is constant and as a result
we return to the direct product case. On the other hand if $f\not=0$, taking the contraction
of (\ref{dfw}) with $\omega_{(2)}$, we find that  $dw=0$ which contradicts
the first relation in (\ref{wtheta7}). Thus $f$ must be zero and so $A$ is constant leading
again to the direct product case. Combining this with our result for the
direct product $AdS_7$ gravitational duals, we have shown that
there are no  $AdS_7$  solutions within  this class.

 \vskip 0.5cm
\noindent{\bf Acknowledgements} \vskip 0.1cm
\noindent  GP thanks the PH-TH Division at CERN for hospitality
where parts of this work were done. UG is supported by the Knut and Alice Wallenberg Foundation. JG is supported by the EPSRC grant, EP/F069774/1.
GP is partially supported by the EPSRC grant EP/F069774/1 and the STFC rolling grant ST/G000/395/1.
\vskip 0.5cm

 \setcounter{section}{0}

 \appendix{From AdS spaces to black hole horizons}

It has been shown in \cite{wald} under certain conditions that the  near horizon geometry of extreme black holes can be written as
\bea
ds^2=2 du (dr+ r h- r^2{\Delta\over2} du)+ ds^2({\cal S})~,
\la{nhm}
\eea
where the 1-form $h$ and function $\Delta$ depend on the coordinates
of the near horizon section ${\cal S}$. In the near horizon calculations, it is
assumed that the horizon section ${\cal S}$ is compact without boundary. Here, we shall not
make such an assumption and allow ${\cal S}$ to be non-compact.
The  metric (\ref{nhm})  includes both the direct  and warped product
metrics  of $AdS$ or Minkowski spaces with a transverse ``internal''
space $X$, respectively. The change of coordinates  that takes these metrics to the near horizon form (\ref{nhm}) is an Eddington-Finkelstein transformation, for $AdS_2$ and $AdS_3$ see also  \cite{lucieti}.

\subsection{Direct products}

To see this, first consider the direct product metric $ds^2(AdS_n\times X)=ds^2(AdS_n)+ds^2(X)$ for $n>2$ and focus on the metric of $AdS_n$. In particular, write
\bea
ds^2(AdS_n)= dz^2+ e^{2 z\over\ell} \big(2 du dv +\sum_k (dx^k)^2\big)~,
\eea
and perform the coordinate transformation
\bea
u=u, z=z, v=e^{-{2z\over\ell}} r
\eea
to find that
\bea
ds^2(AdS_n\times X)&=&ds^2(AdS_n)+ ds^2(X)
\cr
&=&2 du (dr- {2\over\ell} r Z^1)+ (Z^1)^2+ \sum_k (Z^k)^2+ ds^2(X)~,
\la{adsn}
\eea
where we have introduce the frame
\bea
Z^1=dz~,~~~Z^k=e^{ z\over\ell} dx^k~.
\la{zframe}
\eea
Clearly (\ref{adsn}) is of the form of near horizon geometry metrics  for black holes (\ref{nhm})
 with  $\Delta=0$ and  $h=-{2\over\ell} Z^1$.  Furthermore observe that the $Z$ frame satisfies
 the differential relation
 \bea
 dZ^1=0~,~~~dZ^k={1\over\ell} Z^1\wedge Z^k~.
 \la{dzet}
 \eea

The $AdS_2$ case is special. For this write
\bea
ds^2(AdS_2)= {\ell^2\over\rho^2} (-dt^2+d\rho^2)
\eea
and perform the coordinate transformations
\bea
t=u+\ell^2 r^{-1}~,~~~\rho=\ell^2 r^{-1}~,
\eea
to find that the metric $ds^2(AdS_2\times X)=ds^2(AdS_2)+ds^2(X)$ can be written as in (\ref{nhm}) for $h=0$ and $\Delta=\ell^{-2}$.

The direct product metric on $\bR^{n-1,1}\times X$ is clearly a special case of (\ref{nhm})
by taking $h=\Delta=0$. Of course if $n>2$, ${\cal S}$ should be invariant under the
Euclidean group $SO(n-2)\ltimes \bR^{n-2}$ so that the full metric on the product is invariant
under the Poincare group of the $\bR^{n-1,1}$ subspace.
\subsection{Warped products}

Consider the metric
\bea
ds^2= A^2 \big[ e^{2z/\ell} \big(2 dudv+ \sum_{k> 1} (dx^k)^2\big)+dz^2\big]+ ds^2(X)~,
\la{wmetr}
\eea
on the warped product $AdS_n\times_w X$,
where the  function $A$ depends only on the coordinates of $X$. This restriction
is consistent with the requirement that the spacetime metric is invariant under the isometries of $AdS_n$.

First suppose that $n>2$ and consider the coordinate transformation
\bea
v= A^{-2} e^{-2z/\ell} r~,~~~u=u~,~~~z=z~,~~~x^k=x^k~,
\eea
to find that
\bea
ds^2=2 du (dr-{2r\over \ell} Z^1-rd\log A^2)+  A^2\sum_{k\geq 1} (Z^k)^2+ ds^2(X)~,
\la{wpm}
\eea
where the frame $(Z^1, Z^k)$ is defined as in (\ref{zframe}).
Clearly the metrics on warped products of $AdS_n$, $n>2$, spaces are also special cases of near horizon black hole geometries.

It is also easy to see that the warped product metrics on $AdS_2\times_w X$ are special cases of near horizon metrics (\ref{nhm}). Writing the metric on $AdS_2\times_w X$ as
 \bea
 ds^2= 2 A^2 du (dv-\ell^{-2} v^2 du)+ ds^2(X)~,
 \eea
and performing the the coordinate transformation $u=u, r=v A^2$, one can show that it can be
re-expressed as the near horizon metric  (\ref{nhm})
with $\Delta=\ell^{-2} A^{-2}$ and $h=-d\log A^2$.

Warped product metrics of Minkowski spaces can also be written as in (\ref{nhm}). The formulae
of the AdS spaces described above can be adapted to the Minkowski spaces by taking the limit
of large radius $\ell\rightarrow \infty$.

The inclusion of direct and warped products of AdS spaces in the near horizon geometries of black holes has the advantage that all these spaces can be understood in a unified way. In particular, the horizon section  ${\cal S}$  is now spanned by the frame
$(Z^1, Z^k)$ and that of $X$. Therefore for a $AdS_n\times_w X$ spacetime,
\bea
{\cal S}=H_{n-2}\times_w X~,
\eea
 where $H_{n-2}$ is the (n-2)-dimensional hyperbolic space. For all $AdS_n\times_w X$
 the underlying geometry of ${\cal S}$, as  specified by the KSEs, is the same irrespective of the $AdS_n$ subspace but depending
 on the number of supersymmetries preserved. The only difference is that for each $AdS_n\times_w X$ space, one has to consider that  ${\cal S}$ which is  a warped product
  with the appropriate hyperbolic  space.

It is clear that the classification of the local geometries of $AdS\times_w X$ spaces is a special case of that of black hole horizons.  As a paradigm, we shall present the solution of both the field and KSEs for $AdS_n\times_wX$ spaces, for n=3,5,7, in IIB supergravity with only
 5-form flux utilizing the  classification of IIB horizons in \cite{iibhor}. In this context,
  all ${\cal S}=H_{n-2}\times_w X$ spaces are 2-strong Calabi-Yau manifolds with torsion.

\appendix{Heterotic AdS solutions}

AdS backgrounds, and particularly $AdS_3$,  arise naturally in the classification of supersymmetric heterotic backgrounds \cite{hetclas} and have been investigated extensively
in \cite{ohta}. Furthermore,
it is known that heterotic horizons are either products $\bR^{1,1}\times {\cal S}^8$, where
${\cal S}^8$ is a product of Berger manifolds, or  certain fibrations of $AdS_3$ over
 7-dimensional manifolds $X^7$ which admit at least a $G_2$-structure compatible with a
 connection with skew-symmetric torsion. The twisting of
the fibration is with respect to a $U(1)$ connection. The $G_2$ structure appears for solutions preserving 2 supersymmetries and  reduces to $SU(3)$, to $SU(2)$ and to $SU(2)$ for horizons preserving
4, 6 and 8 supersymmetries, respectively.
Thus the geometry of heterotic $AdS_3$ solutions can be read off from that of black hole horizons in \cite{hh}. Because of this, we shall not  proceed to give the details of the various  geometries.  However, as in the IIB case, we shall examine the symmetry of the heterotic $AdS_3$
backgrounds.

The fibration structure of $AdS_3$ over $X^7$ breaks the $SO(2,2)$ isometry group of $AdS_3$ to a subgroup which always contains $SL(2, \bR)$. This can be easily seen in \cite{hh}. In the
heterotic case $SO(2,2)$ is broken by both the metric and the 3-form flux. To restore
the full $SO(2,2)$ symmetry, one has to take the curvature of the fibration to vanish which implies
\bea
dh=0~,
\la{dh0}
\eea
where we follow closely the terminology of \cite{hh}. In such a case, the solution becomes
$AdS_3\times X^7$. If the solution preserves 2 supersymmetries, then $X^7$ is a strong,
conformally balanced manifold which is compatible with a connection with skew-symmetric torsion, i.e.~
\bea
d\tilde H_{(7)}=0~,~~~\theta_\varphi=2d\Phi~,~~~{\rm hol}(\hat\nabla_{(7)})\subseteq G_2~,
\eea
where $H_{(7)}$ is the torsion 3-form on $X^7$, $\theta_\varphi$ is the Lee form of the fundamental $G_2$ form, and $\hat\nabla_{(7)}$ is the connection with skew-symmetric torsion
$H_{(7)}$. In the remaining cases, the description of the geometry for $AdS_3$ backgrounds with
$SO(2,2)$ symmetry  follows from that for the black hole horizons
of \cite{hh} by systematically inserting (\ref{dh0}) into the equations.


\end{document}